\documentclass[prd,aps,preprintnumbers,floatfix,showpacs,preprint,superscriptaddress]{revtex4}
\usepackage{epsfig}
\usepackage{graphics}
\usepackage{latexsym}
\usepackage{amsmath}
\usepackage{amssymb}
\usepackage{rotating}

\newcommand{\mpi}{m_\pi}
\newcommand{\be}{\begin{equation}}
\newcommand{\ee}{\end{equation}}

\begin{document}

\preprint{ADP-05-09/T619}
\preprint{DESY 05-060}

\title{Spin-$\frac{3}{2}$ Pentaquark Resonance Signature in Lattice QCD}

\author{B.~G.~Lasscock}
\affiliation{    Special Research Centre for the
                 Subatomic Structure of Matter,
                 and Department of Physics,
                 University of Adelaide, Adelaide SA 5005,
                 Australia}
\author{D.~B.~Leinweber}
\affiliation{    Special Research Centre for the
                 Subatomic Structure of Matter,
                 and Department of Physics,
                 University of Adelaide, Adelaide SA 5005,
                 Australia}
\author{W.~Melnitchouk}
\affiliation{Jefferson Lab, 12000 Jefferson Ave.,
             Newport News, VA 23606 USA}
\author{A.~W.~Thomas}
\affiliation{    Special Research Centre for the
                 Subatomic Structure of Matter,
                 and Department of Physics,
                 University of Adelaide, Adelaide SA 5005,
                 Australia}
\affiliation{Jefferson Lab, 12000 Jefferson Ave.,
             Newport News, VA 23606 USA}
\author{A.~G.~Williams}
\author{R.~D.~Young}
\affiliation{    Special Research Centre for the
                 Subatomic Structure of Matter,
                 and Department of Physics,
                 University of Adelaide, Adelaide SA 5005,
                 Australia}
\affiliation{Jefferson Lab, 12000 Jefferson Ave.,
             Newport News, VA 23606 USA}
\author{J.~M.~Zanotti}
\affiliation{    Special Research Centre for the
                 Subatomic Structure of Matter,
                 and Department of Physics,
                 University of Adelaide, Adelaide SA 5005,
                 Australia}
\affiliation{	John von Neumann-Institut f\"ur Computing NIC/DESY,
                 15738 Zeuthen, Germany}

\begin{abstract}
The possible discovery of the $\Theta^+$ pentaquark has motivated a
number of studies of its nature using lattice QCD.  While all the
analyses thus far have focused on spin-${1\over 2}$ states, here we
report the results of the first exploratory study in quenched lattice
QCD of pentaquarks with spin ${3 \over 2}$.  For the spin-${3 \over
2}$ interpolating field we use a product of the standard $N$ and $K^*$
operators.  We do not find any evidence for the standard lattice
resonance signature of attraction 
(i.e., binding at  quark masses near the physical regime) 
in the $J^P = {3 \over 2}^-$
channel.  Some evidence of binding is inferred in the isoscalar ${3
\over 2}^+$ channel at several quark masses, in accord with the
standard lattice resonance signature.  This suggests that this is a
good candidate for the further study of pentaquarks on the lattice.
\end{abstract}

\vspace{3mm}

\maketitle

\section{Introduction}

The recent reported observations of the strangeness +1 pentaquark,
$\Theta^+$, having minimal quark content $uudd\bar s$, have led
to a tremendous effort aimed at understanding its properties both
experimentally and theoretically.
Many model studies and phenomenological analyses have explored various
aspects of its structure and production mechanisms, and have at the same
time have revealed a number of challenges for its interpretation as an
exotic resonance with a particularly narrow width (for recent
experimental reviews see
Refs.~\cite{Hicks:2004ge,Hicks:2005pm,Hicks:2005gp,Dzierba:2004db}).

To tackle this problem from first principles in QCD, a number of
lattice studies have recently been undertaken
\cite{Lasscock:2005tt,Csikor:2003ng,Sasaki:2003gi,%
Mathur:2004jr,Ishii:2004qe,Ishii:2004ib,Takahashi:2004sc,%
Takahashi:2005uk,Chiu:2004gg,Alexandrou:2005gc,Csikor:2005xb,Holland:2005yt}.
These have used various local spin-${1\over 2}$ interpolating fields
(either $NK$-type or diquark-diquark-$\bar s$ type), and, in the case
of Ref.~\cite{Csikor:2005xb}, also non-local fields.
Several of these studies have interpreted their results as indicating
the presence of a resonance, while others report signals which are
consistent with $NK$ scattering states.

A major challenge in the lattice studies has been the identification
of a resonance state from the $NK$ scattering states.  Several groups
have sought to distinguish the resonance and scattering states by
comparing the masses at different volumes
\cite{Mathur:2004jr,Alexandrou:2005gc,Takahashi:2004sc,%
Takahashi:2005uk,Csikor:2005xb}.
The volume dependence of the residue of the lowest lying state has
also been proposed as a way to identify the nature of the state
\cite{Mathur:2004jr,Alexandrou:2005gc}.  Alternatively, hybrid
boundary conditions have been used in
Refs.~\cite{Ishii:2004qe,Ishii:2004ib} to differentiate the resonance
in the negative parity channel from the $S$-wave $NK$ scattering
state.

In Ref.~\cite{Lasscock:2005tt} we employed a complimentary approach to
investigate spin-${1\over 2}$ pentaquark resonances by searching for
evidence of sufficient attraction between the constituents of the
pentaquark state such that the resonance mass becomes lower than the
sum of the free decay channel masses. We labeled this pattern as
``the standard lattice resonance signature'' because this signature is
observed for conventional baryon resonances studied on the lattice
\cite{Leinweber:2004it,Melnitchouk:2002eg,Zanotti:2003fx,Sasaki:2001nf,Gockeler:2001db}.
By comparing the masses of the spin-${1\over 2}$
five-quark states to the mass of the decay channel we found no binding 
at any quark mass and hence no evidence for such attraction.
The absence of binding cannot be used to exclude the possibility of
a resonance, as the attractive forces simply may not be strong enough to
provide binding. On the other hand, the presence of binding would provide
a compelling resonance signature warranting further study.

One of the major puzzles in pentaquark phenomenology has been the
anomalously small width ($\alt 1$~MeV) observed in the experiments
which have produced a positive signal.
A possible explanation for this may be that if the pentaquark has
$J^P={3\over 2}^-$, its decay to $N+K$ must be via a $D$-wave,
which would consequently be suppressed.
In this paper we therefore extend the analysis of
Ref.~\cite{Lasscock:2005tt} to spin-${3\over 2}$ pentaquarks.
We examine both the positive and negative parity states, in both
the isoscalar and isovector channels.
In Sec.~II we describe the interpolating field and outline the
lattice techniques employed in this analysis.
The results are presented in Sec.~III, where we discuss in detail
the mass splittings between the pentaquark and two-particle
scattering states.
Finally, conclusions and suggestions for future work are summarised
in Sec.~IV.

\section{Lattice Details}

\subsection{Interpolating fields}

The simplest $NK$-type interpolating field used in lattice simulations,
referred to in Ref.~\cite{Lasscock:2005tt} as the ``colour-singlet''
$NK$ field, has the form:
\begin{equation}
\label{eq:NK:sing}
\chi_{NK} = {1 \over \sqrt{2}} \epsilon^{abc}
	(u^{T a} C \gamma_5 d^b)
	\left\{ u^c (\bar s^e i \gamma_5 d^e)\
		\mp\ (u \leftrightarrow d)
	\right\}\ ,
\end{equation}
where the $-$ and $+$ corresponds to the isospin $I=0$ and $1$ channels, 
respectively.
This field has spin $\frac{1}{2}$, and transforms negatively under
the parity transformation $q \to \gamma_0 \, q$.

One can access spin-$\frac{3}{2}$ states by replacing the spin-0
$K$-meson part of $\chi_{NK}$ with a spin-1 $K^*$ vector meson
operator,
\begin{equation}
\label{eq:NK*:sing}
\chi^{\mu}_{NK^{*}} = {1 \over \sqrt{2}} \epsilon^{abc}
	(u^{T a} C \gamma_5 d^b)
	\left\{ u^c (\bar s^e i \gamma^{\mu} d^e)\
		\mp\ (u \leftrightarrow d)
	\right\}\ ,
\end{equation}
where again the $-$ and $+$ corresponds to the isospin $I=0$ and $1$
channels, respectively.
The field $\chi^{\mu}_{NK^*}$ transforms as a vector under the
parity transformation, and
has overlap with both spin-$\frac{1}{2}$ and spin-$\frac{3}{2}$
pentaquark states.
States of definite spin can be projected from $\chi^{\mu}_{NK^*}$ by
applying appropriate projectors, as discussed in the next section.

\subsection{Lattice Techniques}

The masses of the spin-$\frac{1}{2}$ and spin-$\frac{3}{2}$ pentaquark
states are obtained from the two-point correlation function
\begin{eqnarray}
\label{eq:2-pt}
\mathcal{G}_{\mu\nu}(t,{\vec p})
&=& \sum_{\vec x}\ \exp({-i {\vec p} \cdot {\vec x}}) 
\left\langle 0 \left|
   T\ \chi_{\mu}(x)\ \bar\chi_{\nu}(0)\
\right| 0 \right\rangle\ .
\end{eqnarray}
To project states of definite spin from the correlation function
$\mathcal{G}_{\mu\nu}(t,{\vec p})$ we apply the spin projection
operators \cite{Zanotti:2003fx}
\begin{eqnarray}
P^{\frac{3}{2}}_{\mu\nu}(p)
&=& g_{\mu\nu} - \frac{1}{3}\gamma_{\mu}\gamma_{\nu}
 - \frac{1}{3p^{2}}(\gamma\cdot p \gamma_{\mu}p_{\nu}
		   + p_{\mu}\gamma_{\nu}\gamma\cdot p)\ , \cr
P^{\frac{1}{2}}_{\mu\nu}(p)
&=& g_{\mu\nu} - P^{\frac{3}{2}}_{\mu\nu}(p)\ ,
\end{eqnarray}
for spin-${3\over 2}$ and ${1\over 2}$, respectively.

The spin-projected correlation function receives contributions from
both positive and negative parity states.
The use of fixed boundary conditions in the time direction enables
states of definite parity to be projected using the matrix
\cite{Lee:1998cx,Melnitchouk:2002eg}
\begin{eqnarray}
\label{eq:pProjOp}
\Gamma^{\mp}
= {1 \over 2}
  \left( 1 \pm {M_{B^\pm} \over E_{B^\pm}} \gamma_4 \right)\ ,
\end{eqnarray}
for negative and positive parities, respectively.  We note that this
differs from that of Ref.~\cite{Zanotti:2003fx}, where interpolating
fields transforming as pseudovectors, in accord with the
Rarita-Schwinger spinor-vectors, were used.
Masses of states with definite spin and parity can then be obtained
from the spinor trace of the spin and parity projected correlation
functions,
\begin{eqnarray}
G(t,{\vec p})
&=& {\rm tr_{sp}}
    \left[ \Gamma \, \mathcal{G}_{\mu\nu}(t,{\vec p})\, P^{\nu \mu}(p) 
    \right]						\cr
&=& \sum_B \lambda^B \bar{\lambda}^B
    \exp{(-E_B t)}                                      \cr
&\stackrel{t\to\infty}{=}& \lambda^0 \bar{\lambda}^0 \exp{(-m_0 t)}\ .
\end{eqnarray}
This function is a sum over all states, $B$, with energy $E_B$,
and $\bar{\lambda}^B$ and $\lambda^B$ are the couplings of the
state $B$ to the interpolating fields at the source and sink,
respectively.
Note that as in Ref.~\cite{Zanotti:2003fx}, we consider the case $\mu = 3$ to reduce the computational 
cost of the calculation. 

Since the contributions to the two-point function are exponentially
suppressed at a rate proportional to the energy of the state,
at zero momentum the mass of the lightest state, $m_0$, is obtained
by fitting a constant to the effective mass,
\begin{eqnarray}
M^{{\rm eff}}(t)
&=& \ln{ \left( G(t,\vec 0) \over G(t+1,\vec 0) \right)}  \cr
&\stackrel{t\to\infty}{=}& m_0 \ .
\end{eqnarray}
Following our previous work \cite{Lasscock:2005tt}, we search for evidence that the 
resonance mass has become smaller than the sum of the free decay channel mass for
pentaquark states created by the interpolating field $\chi^{\mu}_{NK^{*}}$.
For this purpose it is useful to 
define an effective mass splitting.  For example, in an $S$-wave decay
channel,
\begin{eqnarray}
\Delta M^{{\rm eff}}(t)
&\equiv& M_{5q}^{\rm eff}(t)
- \ ( M_B^{\rm eff}(t) + M_M^{\rm eff}(t) )		\cr
&\stackrel{t\to\infty}{=}& m_{5q} - ( m_B +  m_M )\ ,
\end{eqnarray}
where $M_B^{\rm eff}(t)$ and $M_M^{\rm eff}(t)$ are the appropriate
baryon and meson effective masses for a specific channel.  For a
$P$-wave decay channel, the effective masses are combined with the
minimum nontrivial momentum on the lattice, $2\pi/L$, to create the
effective energy, 
$E^{{\rm eff}}(t) = \sqrt{ ( M^{{\rm eff}}(t) )^{2} + (2\pi/L)^{2} }$, for each decay particle,
where $L$ is the lattice spatial extent.
The advantage of this technique is that it measures a correlated
mass difference, thereby suppressing the sensitivity to systematic
uncertainties (such as using different fitting ranges).  Moreover,
correlations in the effective masses can cancel, leading to a more
accurate determination of the mass splitting.

\begin{figure}[t]
\includegraphics[height=9.0cm,angle=90]{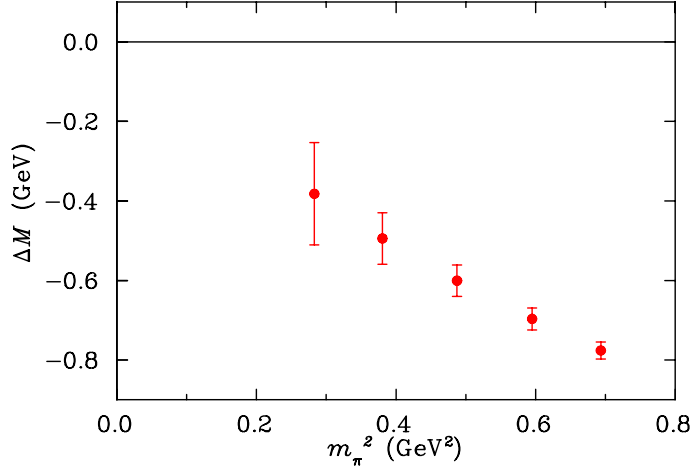}
\caption{\label{fig:Delta}
	Mass difference between the
	$I(J^P)={3 \over 2}({3 \over 2}^+)\ \Delta(1232)$
	and the $P$-wave $N+\pi$ decay channel. }
\end{figure}

As an example of a lattice signature of a resonance, we consider the
mass splitting between the $J^P=3/2^+$ $\Delta$ baryon and the energy
of the $P$-wave $N+\pi$ decay channel.
In Fig.~\ref{fig:Delta} the typical resonance signature is clearly seen,
where the $\Delta$ is bound on the lattice 
at heavier than physical quark masses,
because of its lower energy
compared with the free decay channel.

For a pentaquark resonance we shall apply the same criteria, and
consider the mass splittings between the pentaquark state and the
corresponding baryon and meson free two-particle scattering states.
The signal we are searching for is evidence of a pentaquark bound state
at quark masses near the physical regime.
In Table~\ref{tab:dchannel} we summarise the lowest energy decay
channels for the various isospin, spin and parity quantum numbers
considered in this analysis.

\begin{table}
\caption{Lowest energy decay channels for each pentaquark state
	on the lattice, where the $\Delta$ baryon is bound.
\label{tab:dchannel}}
\begin{tabular}{cl}
\hline
$I(J^{P})$             & Decay channel     \cr
\hline
\hline
$0,1(\frac{1}{2}^{-})$  & $S$-wave $N+K$      \cr
$0,1(\frac{1}{2}^{+})$  & $P$-wave $N+K$      \cr       
$0(\frac{3}{2}^{-})$    & $S$-wave $N+K^{*}$  \cr
$0(\frac{3}{2}^{+})$    & $P$-wave $N+K$      \cr
$1(\frac{3}{2}^{-})$    & $S$-wave $\Delta+K$ \cr
$1(\frac{3}{2}^{+})$    & $P$-wave $N+K$  \cr
\hline
\end{tabular}
\end{table}

In the case of the $\Delta$ baryon, the binding is seen to become stronger at
larger quark masses.  Indeed, from their minimal quark content in the
heavy quark limit, one expects to recover a $\Delta$ to $N\, \pi$ mass
ratio of 3/5.  In the case of a pentaquark resonance, the analogous
mass ratio will be 1 in the heavy quark limit and 
the mass splitting will vanish relative to the hadronic mass scale in the heavy quark
limit.  Hence a lattice resonance signature for a pentaquark state is
binding (a negative mass splitting) at intermediate quark masses, above the physical regime,
with a general trend of binding as a fraction of hadron mass towards zero as the heavy quark limit is
approached.

\subsection{Lattice Simulation Formalism}

This analysis is based on the ensemble of 290, $20^{3}\times 40$ $SU(3)$ gauge-field
configurations considered in \cite{Lasscock:2005tt}. 
Using the mean-field
${\cal O}(a^2)$-improved Luscher-Weisz plaquette plus rectangle action
\cite{Luscher:1984xn},  the gauge configurations are generated via
the Cabibbo-Marinari pseudoheat-bath algorithm with three diagonal
SU(2) subgroups looped over twice.  The simulations are performed
using a parallel algorithm with appropriate link partitioning, as
described in Ref.~\cite{Bonnet:2000db}. The lattice spacing is 0.128(2)
fm, determined using the Sommer scale $r_{0}=0.49$ fm.

For the fermion propagators, we use the FLIC fermion action
\cite{Zanotti:2001yb}, an ${\cal O}(a)$-improved fermion action with
excellent scaling properties providing near continuum results at
finite lattice spacing \cite{Zanotti:2004dr}.

A fixed boundary condition in the time direction is implemented by
setting $U_t(\vec x, N_t) = 0\ \forall\ \vec x$ in the hopping terms
of the fermion action.  Periodic boundary conditions are imposed in
the spatial directions.  To explore the effects of the fixed boundary
condition we have examined the effective mass of the pion correlation
function and the associated $\chi^{2}_{{\rm dof}}$ obtained in various
fits.  The pion is selected as it has the longest correlation length
and will be a worst case scenario for the boundary effects.  We find
that the fixed boundary effects are completely negligible prior to
time slice 30, which is the limit of signal in the pentaquark
correlation functions presented below.

Gauge-invariant Gaussian smearing \cite{Gusken:1989qx} in
the spatial dimensions is applied at the fermion source at $t=8$ to
increase the overlap of the interpolating operators with the ground
states. Six quark masses are used in the calculations, with $\kappa =
\{ 0.12780, 0.12830, 0.12885, 0.12940, 0.12990, 0.13025 \}$ providing
$a m_\pi = \{0.540,0.500,0.453,0.400,0.345,0.300 \}$ 
\cite{Boinepalli:2004fz}.
The strange quark mass is taken to
be the third largest ($\kappa = 0.12885$) quark mass. 
This $\kappa$ provides a pseudoscalar mass of $697$~MeV which compares
well with the experimental value of
$\sqrt{2M_{{\rm K}}^{2} - M_{\pi}^{2}} = 693$~MeV motivated by leading order
chiral perturbation theory. The error analysis is performed by a
second-order, single-elimination jackknife, with the $\chi^2$ per degree
of freedom obtained via covariance matrix fits. Further details of
the fermion action and simulation parameters are provided in
Refs.~\cite{Zanotti:2001yb,Zanotti:2004dr} and \cite{Lasscock:2005tt}
respectively.

\section{ Results }

In this section we present our results for the masses of
spin-${3 \over 2}$ pentaquarks for both negative and positive
parity, in both the isoscalar and isovector channels.

\subsection{Negative parity isoscalar channel}

\begin{figure}[tp]
\includegraphics[height=9.0cm,angle=90]{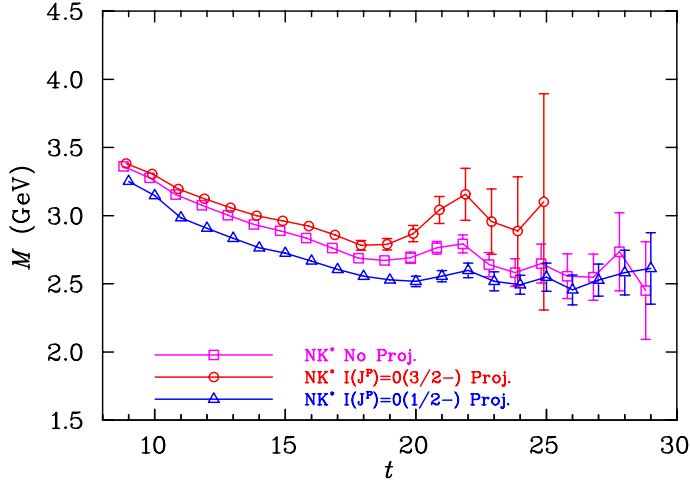} 
\caption{\label{fig:I0.neg_diff} Effective masses of the negative
	parity projected, isoscalar correlation functions calculated
	with the $NK^{*}$ pentaquark interpolator, $\chi_{NK^*}^\mu$.
	The mass corresponding to the unprojected $G_{33}$ correlation
	function (squares) is compared with that of the
	spin-$\frac{1}{2}$ (triangles) and spin-$\frac{3}{2}$ (circles)
	projected correlation functions.
	The data corresponds to our heaviest quark mass. }
\end{figure}

We begin the discussion of our results with the isoscalar, negative
parity channel. 
The effect of the spin projection on the correlation function is
highlighted in Fig.~\ref{fig:I0.neg_diff}.  This figure shows the
effective mass plot of the $G_{33} = {\rm tr_{sp}} \{\Gamma^-
\mathcal{G}_{33} \}$ component of the correlation function, which
contains a superposition of spin-$\frac{1}{2}$ and spin-$\frac{3}{2}$
contributions.
Upon spin projection, as described by Eq.~(\ref{eq:pProjOp}), two
distinct states are identified.

\begin{figure}[tp]
\includegraphics[height=9.0cm,angle=90]{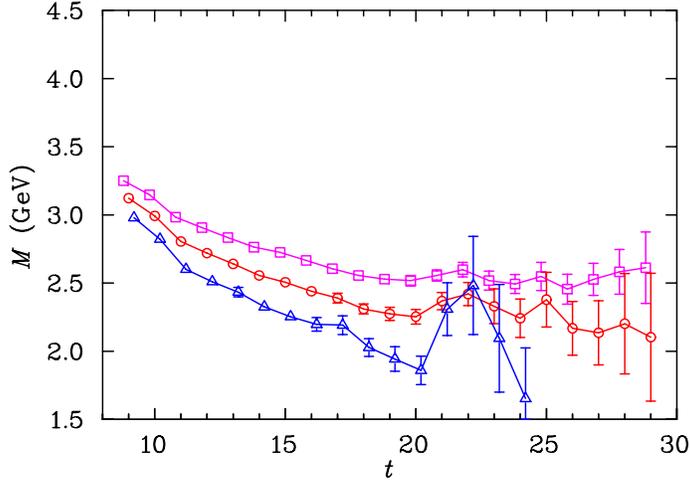} 
\caption{\label{fig:I0.neg_12} Effective mass of the
	$I(J^P)=0(\frac{1}{2}^-)$ pentaquark calculated with the
	$NK^{*}$ pentaquark interpolator.  
	The data correspond to $\mpi\simeq 830{\,{\rm MeV}}$ (squares),
	$700{\,{\rm MeV}}$ (circles), and  $530{\,{\rm MeV}}$ (triangles).}
\end{figure}

\begin{figure}[tp]
\includegraphics[height=9.0cm,angle=90]{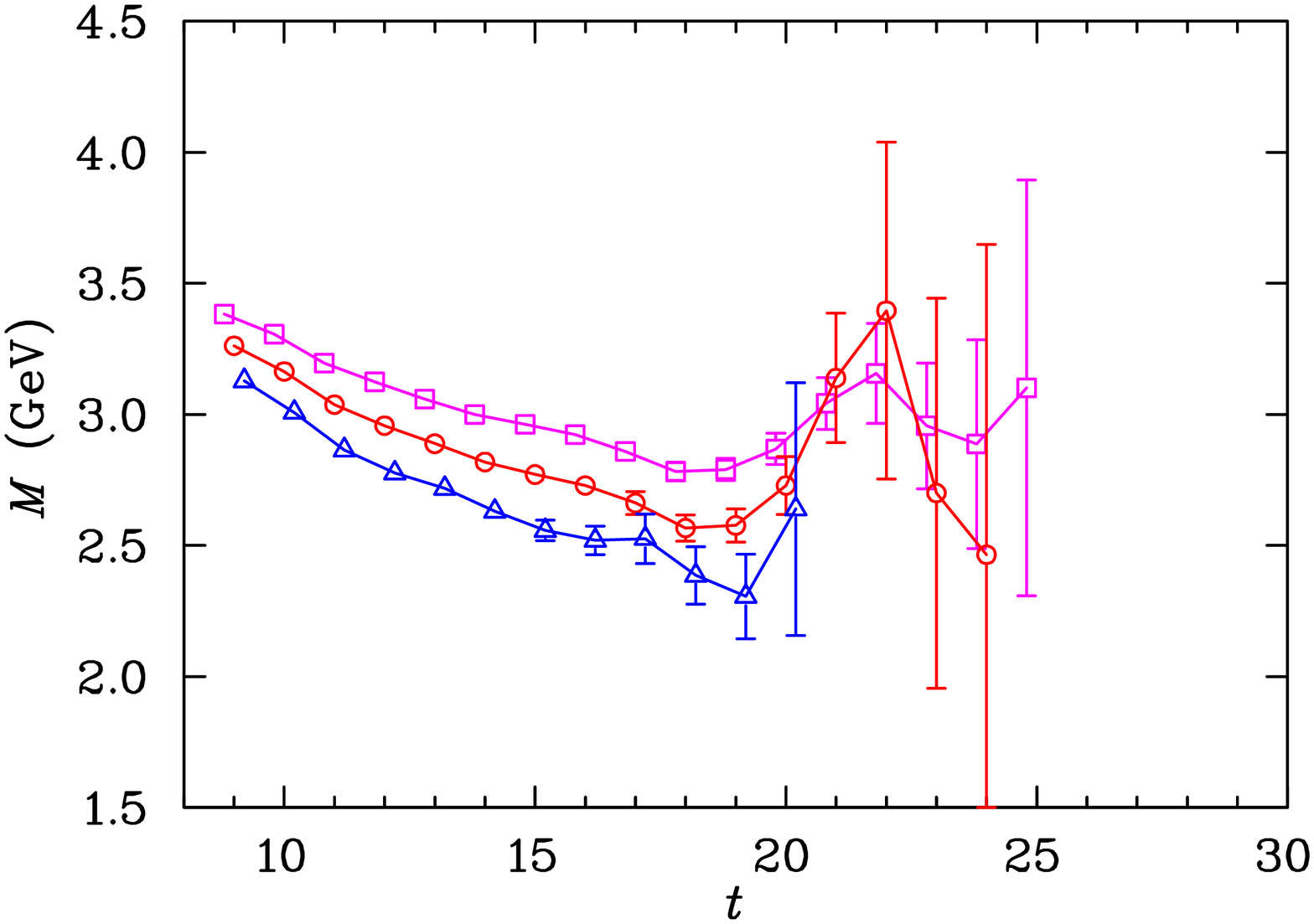} 
\caption{\label{fig:I0.neg_32} As in Fig.~\ref{fig:I0.neg_12}, but for
	the $I(J^P)=0(\frac{3}{2}^-)$ state.}
\end{figure}

\begin{figure}[tp]
\includegraphics[height=9.0cm,angle=90]{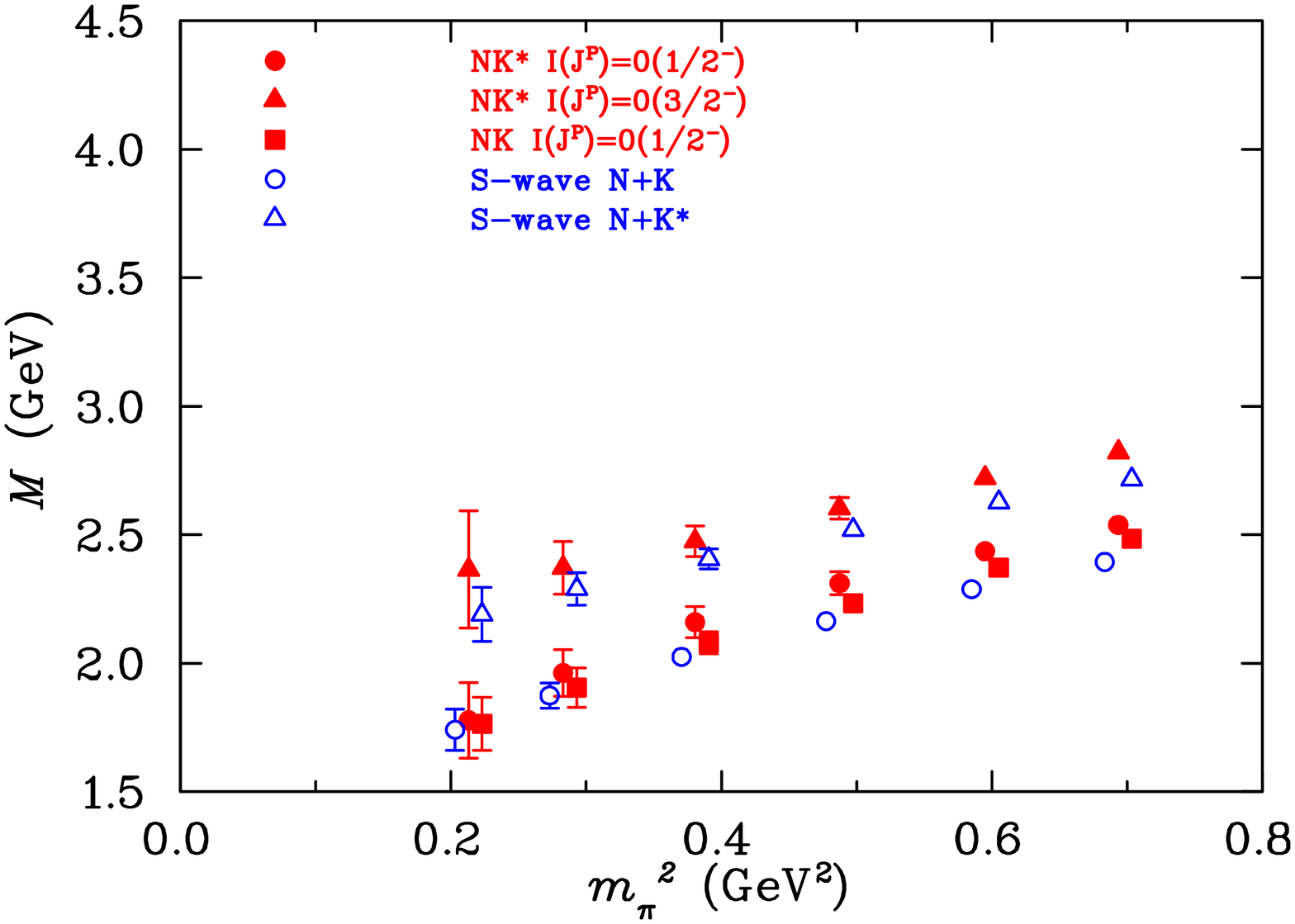} 
\caption{\label{fig:I0.neg} Masses of the $I(J^P)=0(\frac{1}{2}^-)$
	and $0(\frac{3}{2}^-)$ states extracted with the
	$NK^{*}$ interpolating field as a function of $m_\pi^2$.
	For comparison, we also show the mass of the
	$I(J^P)=0(\frac{1}{2}^-)$ state extracted from the $NK$
	pentaquark interpolator \cite{Lasscock:2005tt}, and the masses
	of the $S$-wave $N+K$ and $N+K^{*}$ two-particle states.
	Some of the points have been horizontally offset for clarity.}
\end{figure}

The effective masses of the spin-$\frac{1}{2}$ and spin-$\frac{3}{2}$
states are presented in Figs.~\ref{fig:I0.neg_12} and
\ref{fig:I0.neg_32}, respectively.  To extract the masses of the
lowest energy states from these effective masses, we fit over the time
slices $20-30$ for the spin-$\frac{1}{2}$ and $18-21$ for the
spin-$\frac{3}{2}$ states, respectively, where these intervals have
been selected so as to obtain acceptable values of the
covariance-matrix based $\chi^2$ per degree of freedom
($\chi^2_{\rm{dof}}$), which we restrict to $\chi^{2}_{\rm{dof}}<1.5$.
The resulting masses are presented in Fig.~\ref{fig:I0.neg}, together
with the mass of the $I(J^{P})=0(\frac{1}{2}^{-})$ state extracted
with the standard $NK$ pentaquark operator \cite{Lasscock:2005tt} of
Eq.~(\ref{eq:NK:sing}), and the relevant (non-interacting)
two-particle states.  We obtain the expected result that the masses of
the $I(J^{P})=0(\frac{1}{2}^{-})$ state extracted with the $NK$ and
$NK^{*}$ interpolators are in excellent agreement. The mass of the
$I(J^{P})=0(\frac{3}{2}^{-})$ state is also similar to that of the
$N+K^{*}$ two-particle state, but lies consistently above the latter,
suggesting the presence of some repulsion in this channel.  Therefore
we cannot conclude any evidence of a bound state in this channel.

\subsection{Positive parity isoscalar channel}

\begin{figure}[tp]
\includegraphics[height=9.0cm,angle=90]{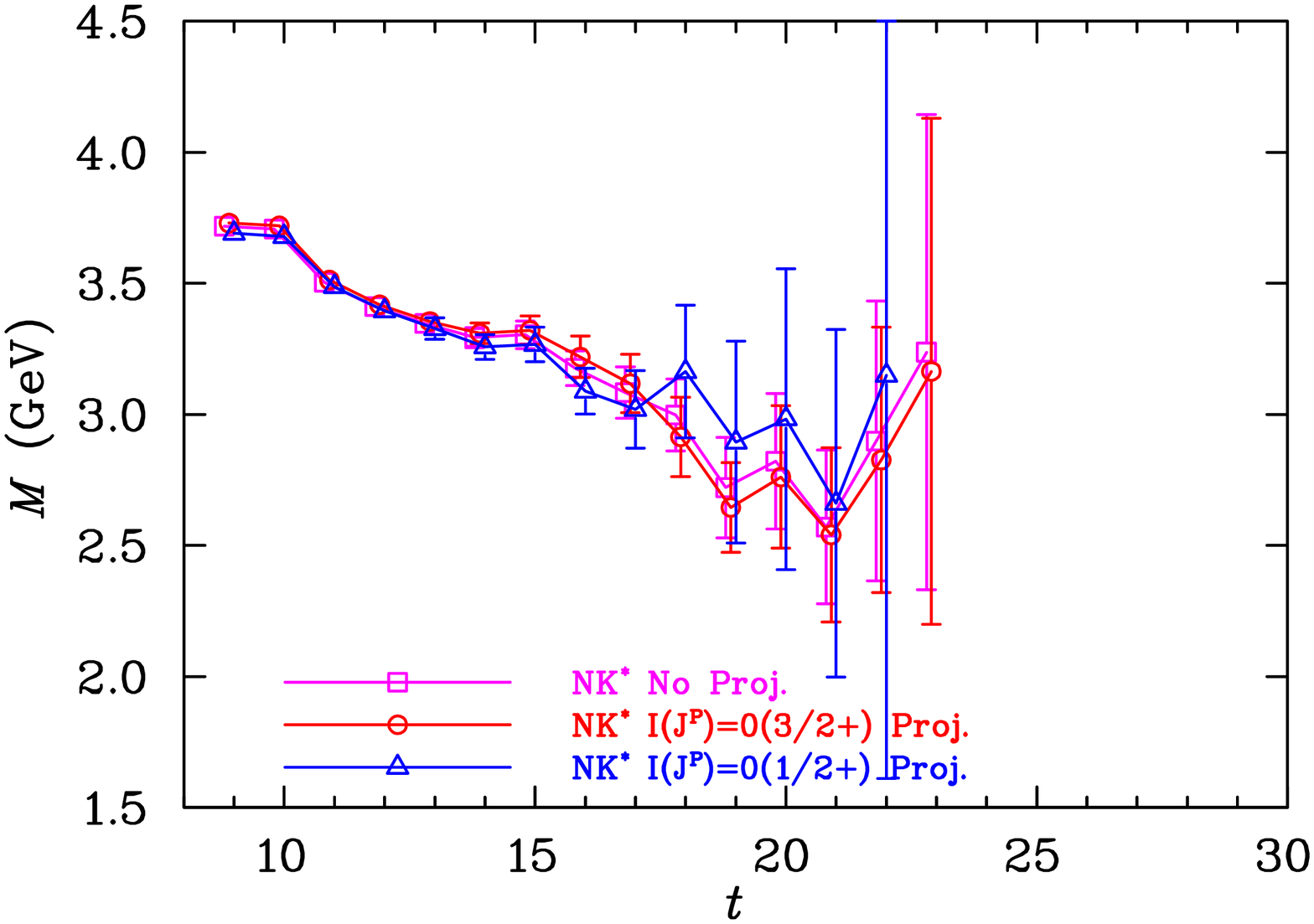} 
\caption{\label{fig:I0.pos_diff} 
        As in Fig.~\ref{fig:I0.neg_diff}, but for the isoscalar
        positive parity channel. }
\end{figure}

\begin{figure}[tp]
\includegraphics[height=9.0cm,angle=90]{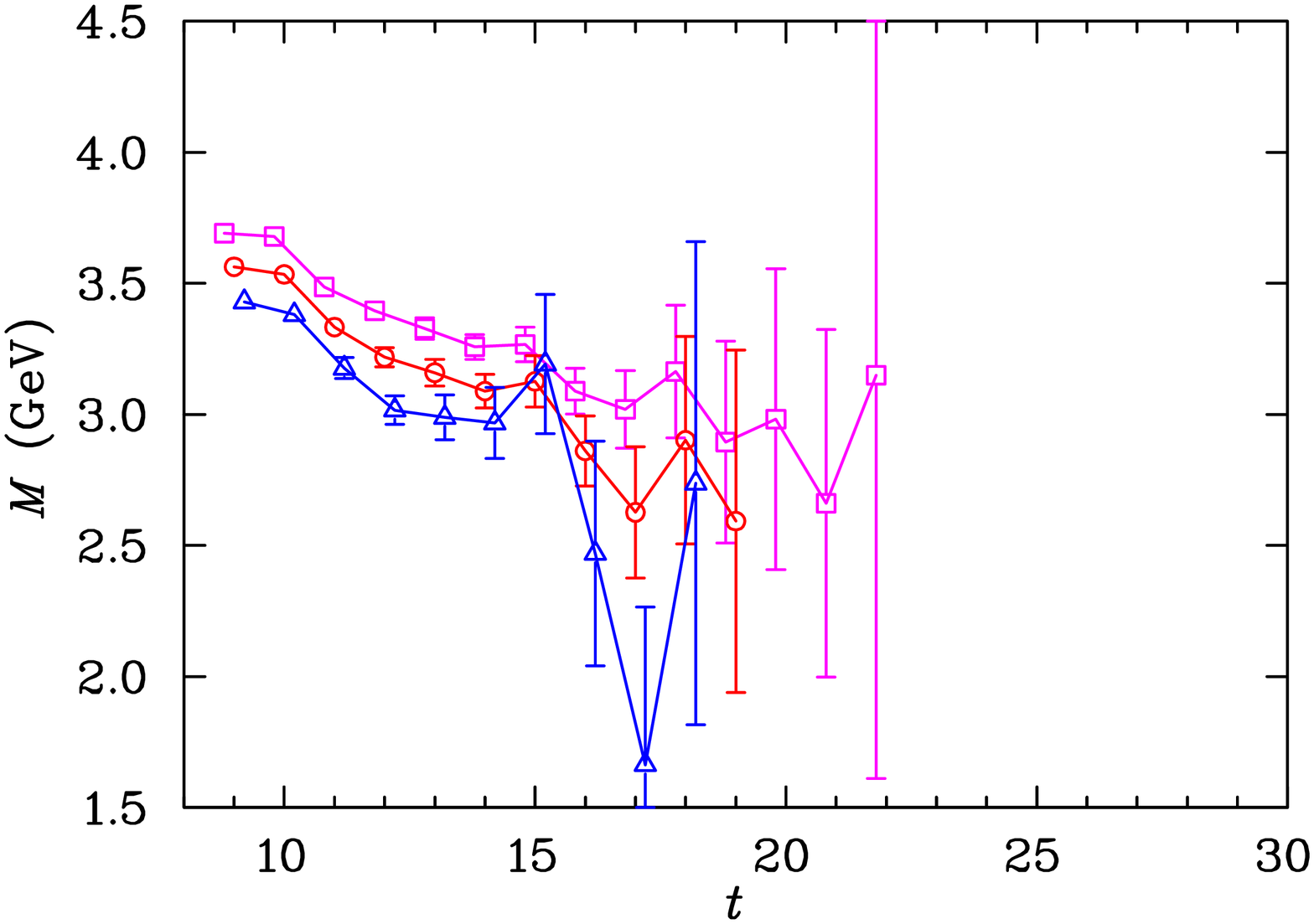} 
\caption{\label{fig:I0.pos_12}
	Effective mass of the $I(J^P)=0(\frac{1}{2}^+)$ pentaquark
	obtained from the $NK^{*}$ interpolating field.  
	The data correspond to $\mpi\simeq 830{\,{\rm MeV}}$ (squares),
	$700{\,{\rm MeV}}$ (circles), and  $530{\,{\rm MeV}}$ (triangles).}
\end{figure}

\begin{figure}[tp]
\includegraphics[height=9.0cm,angle=90]{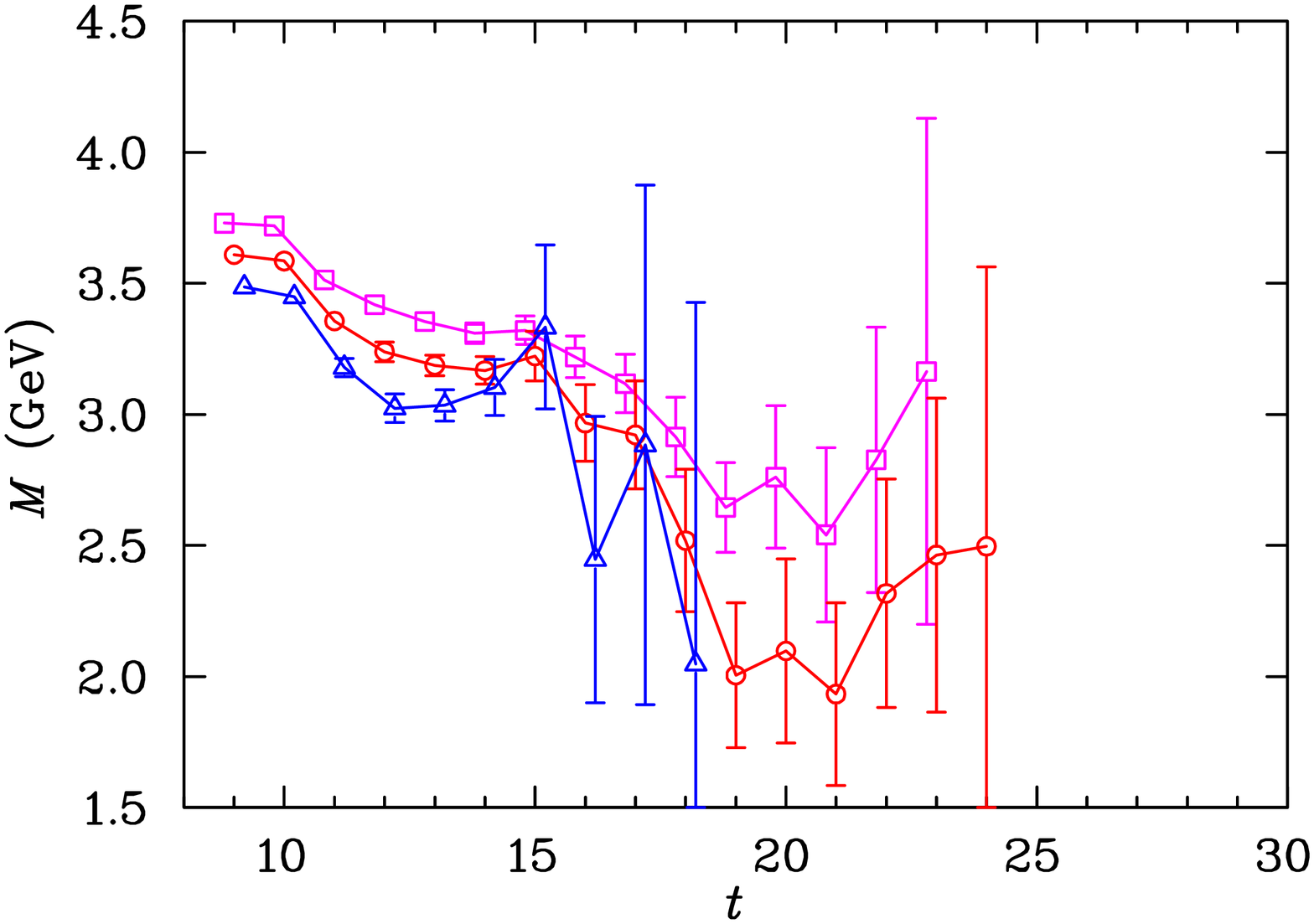} 
\caption{\label{fig:I0.pos_32}
	As in Fig.~\ref{fig:I0.pos_12}, but for the
	$I(J^P)=0(\frac{3}{2}^+)$ state.}
\end{figure}

Next we consider the isoscalar state in the positive parity
channel. Contrary to the negative parity signal, the spin projection
shown in Fig.~\ref{fig:I0.pos_diff} has a less pronounced effect on
the effective masses.  
The effective masses of the spin-$\frac{1}{2}$ and $\frac{3}{2}$ states
are presented in Figs.~\ref{fig:I0.pos_12} and \ref{fig:I0.pos_32},
respectively. The quality of the signal in the positive parity sector
is significantly reduced relative to the negative parity channel, as
in the spin-$\frac{1}{2}$ analysis in Ref.~\cite{Lasscock:2005tt}.

\begin{figure}[tp]
\includegraphics[height=9.0cm,angle=90]{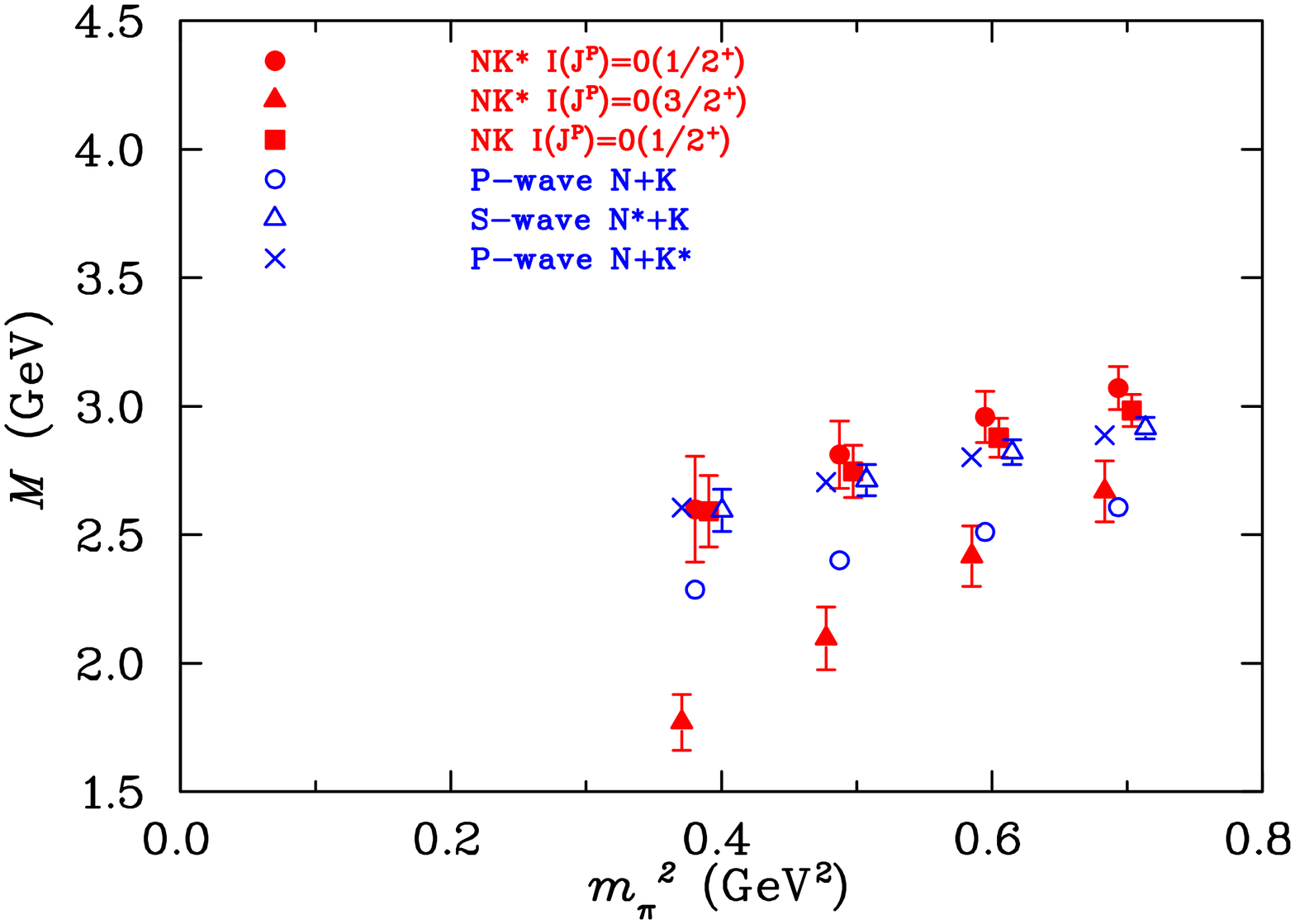} 
\caption{\label{fig:I0_12.pos}
	Masses of the $I(J^P)=0(\frac{1}{2}^+)$ and $0(\frac{3}{2}^+)$
	states determined from the $NK^{*}$ interpolating field as a
	function of $m_\pi^2$.
	For comparison, we also show the mass of the
	$I(J^P)=0(\frac{1}{2}^+)$ state extracted with the $NK$
	pentaquark interpolator \cite{Lasscock:2005tt}, and the masses
	of the $P$-wave $N+K$ and $N+K^{*}$ and $S$-wave $N^{*}+K$
	two-particle states.
	Some of the points have been offset horizontally for clarity.}
\end{figure}

The effective mass of the spin-$\frac{1}{2}$ state is fit at time
slices $18-20$ and the spin-$\frac{3}{2}$ state at time slices
$19-24$.  The poor quality of the signal limits the analysis to
the four largest quark masses considered.  In Fig.~\ref{fig:I0_12.pos}
we show the fitted masses of the two spin states extracted with the
$NK^{*}$ interpolator. For comparison, we display the mass of the
spin-$\frac{1}{2}$ state extracted with the $NK$ interpolator, and
the energies of the relevant two-particle states.  Once
again, we see excellent agreement between the masses of the
spin-$\frac{1}{2}$ states extracted with the $NK$ and $NK^{*}$
interpolators. 

Interestingly, the mass of the spin-$\frac{3}{2}$ state becomes {\em
smaller} than the non-interacting two-particle energy of the $P$-wave
$N+K$ state for intermediate quark masses, i.e., we observe binding.  As discussed in the
previous section and in Ref.~\cite{Lasscock:2005tt}, the transition of
a resonance to a state which lies below the free particle decay
channel at quark masses near the physical quark masses
is the standard resonance signature in lattice QCD.

Moreover the approach to the heavy quark limit is in accord with
expectations.  Recall that in the case of the $\Theta^{+}$, which has
a ``fall-apart'' decay mechanism, quark counting indicates the
$\Theta^{+}$ to $N + K$ mass ratio will approach 1 as the heavy quark
limit is approached.  

At intermediate quark masses, however, one expects the resonance
signature analogous to the $\Delta$ baryon in Fig.~\ref{fig:Delta},
and at the two smallest quark masses shown  the pentaquark lies
below the scattering state, which is the necessary condition for the
presence of binding.  As this result presents the possible existence
of a pentaquark resonance in the physical quark mass regime, it is essential to consider a mass
splitting analysis of the effective masses.

\begin{figure}[tp]
\includegraphics[height=9.0cm,angle=90]{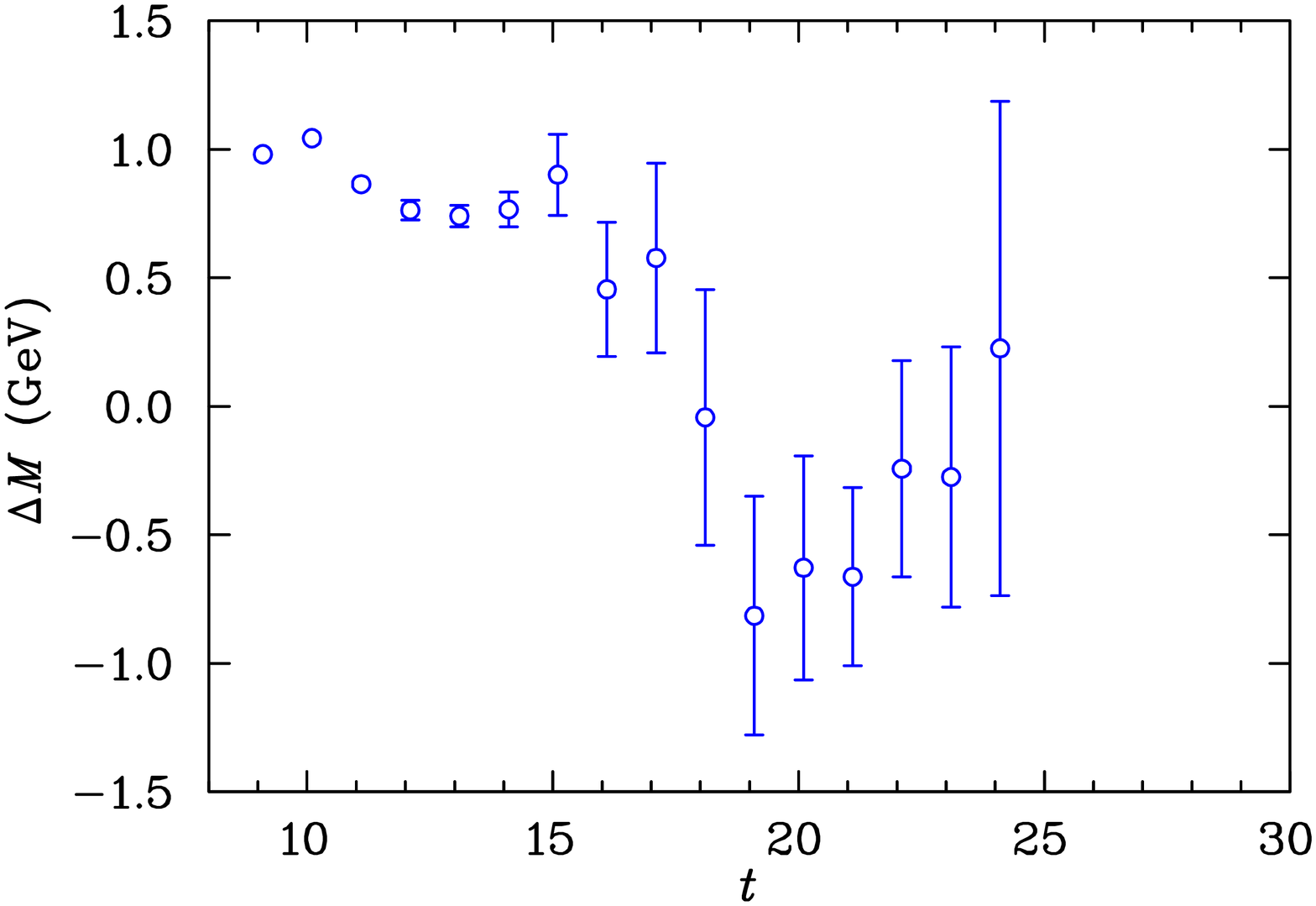} 
\caption{\label{fig:I0_12.pos.s.k4}
         Effective mass splitting between the
	 $I(J^P)=0(\frac{3}{2}^+)$ state extracted with the $NK^{*}$
	 interpolator and the energy of the $P$-wave $N+K$
	 two-particle state for the lightest quark mass shown in
	 Fig.~\ref{fig:I0_12.pos}.  }
\end{figure}

As an indicative example, we review the analysis of the lightest quark
mass presented in Fig.~\ref{fig:I0_12.pos}.  In
Fig.~\ref{fig:I0_12.pos.s.k4} we present the effective mass splitting
between the five-quark state and the lightest non-interacting
two-particle state having the same quantum numbers.

\begin{figure}[tp]
\includegraphics[height=9.0cm,angle=90]{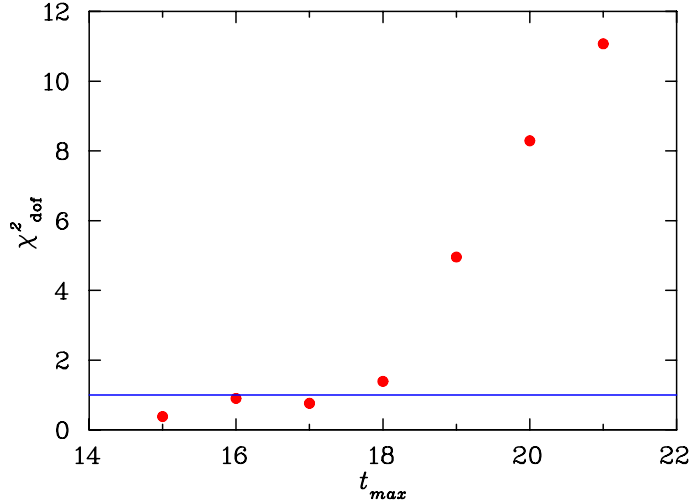} 
\caption{\label{fig:I0_12.pos.s.k4.tmax.chisq}
           The $\chi^{2}_{\rm dof}$ for a series of possible fits
           with a lower bound fixed at time slice $12$ and an upper
           bound shown on the horizontal axis.  }
\end{figure}

At first sight it might be tempting to consider fits early in the
Euclidean time evolution where the errors are small and a possible
plateau catches the eye.  However, it is easy to demonstrate that such
a fit does not describe the lowest lying state in this correlation
function.  For example, Fig.~\ref{fig:I0_12.pos.s.k4.tmax.chisq}
reports the $\chi^{2}_{{\rm dof}}$ for a selection of fits to the
effective mass shown in Fig.~\ref{fig:I0_12.pos.s.k4}, where the lower
bound of the fit window is fixed at $t = 12$ (four time steps from the
source at $t = 8$), and the upper bound of the fit window is plotted on the
horizontal axis.  As soon as time slice 19 is included in the fit the
$\chi^{2}_{{\rm dof}}$ becomes very large and continues to increase with
the inclusion of time slices 20 and 21.  One must therefore conclude
that the results for time slices 19 through 21 are signal rather than
noise, reflecting the true ground state of the correlator.
The effective mass splitting with a ``double plateau'' as in Fig.~\ref{fig:I0_12.pos.s.k4}
can occur when the interpolating fields couple strongly to a more massive state and
relatively weakly to the ground state. The former dominates at early Euclidean times and
the latter at later times. 

To determine an appropriate lower bound for the fit, we begin by
returning to Fig.~\ref{fig:I0_12.pos.s.k4}.  The fluctuation at $t=25$
suggests that there is valid data up to time slice $24$, after which
noise begins to hide the signal.  Indeed, the $\chi^2_{\rm dof}$ is
approximately invariant for fits of $t=19$ through 24 and beyond.
Therefore, time slice $24$ is selected for the upper bound of the fit
window.  

\begin{figure}[tp]
\includegraphics[height=9.0cm,angle=90]{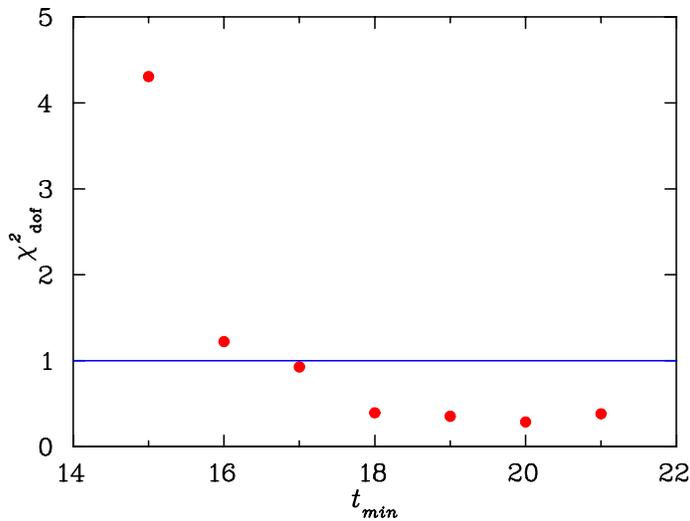} 
\caption{\label{fig:I0_12.pos.s.k4.chisq}
           The $\chi^{2}_{\rm dof}$ for a series of possible fits
           with an upper bound fixed at time slice $24$ and a lower
           bound shown on the horizontal axis.  }
\end{figure}

\begin{figure}[tp]
\includegraphics[height=9.0cm,angle=90]{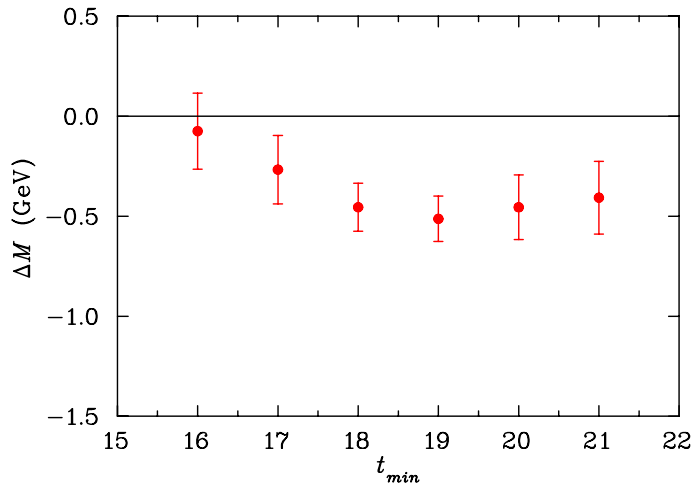} 
\caption{\label{fig:I0_12.pos.s.k4.mass}
           Mass splitting extracted from a series of possible fits
           with an upper bound fixed at time slice $24$ and a lower
           bound shown on the horizontal axis.  }
\end{figure}

The lower bound of the fit interval must be selected with regard to
the systematic time dependence of the effective mass and the
$\chi^{2}_{{\rm dof}}$ of the fit.
Figure~\ref{fig:I0_12.pos.s.k4.chisq} reports the latter criteria,
illustrating the $\chi^{2}_{{\rm dof}}$ for a selection of fits to the
effective mass shown in Fig.~\ref{fig:I0_12.pos.s.k4}, where the lower
bound of the fit window is shown on the horizontal axis and the upper
bound of the fit window is fixed at time slice $24$.
The mass splittings extracted for these fits are shown in
Fig.~\ref{fig:I0_12.pos.s.k4.mass}.
Again, the lower bound of the fit window is shown on the
horizontal axis and the upper bound of the fit window is fixed at time
slice $24$.  The most statistically precise estimate of the mass
splitting is obtained for $t_{\rm min} = 19$, and this result agrees
with all other determinations at the 1$\sigma$ level, with the
exception of $t_{\rm min} = 16$.\par

Having determined that there is genuine signal in the effective mass
splitting at time slice 19, and given the systematic drift in the
results approaching time slice 19 as illustrated in both
Figs.~\ref{fig:I0_12.pos.s.k4} and \ref{fig:I0_12.pos.s.k4.mass}, one
has to conclude that nontrivial contributions from more massive
excited states are still present at time slice 16.  The more
conservative evaluation is to delay the fit to later Euclidean times
where the $\chi^{2}_{{\rm dof}} \alt 1$.

\begin{figure}[tp]
\includegraphics[height=9.0cm,angle=90]{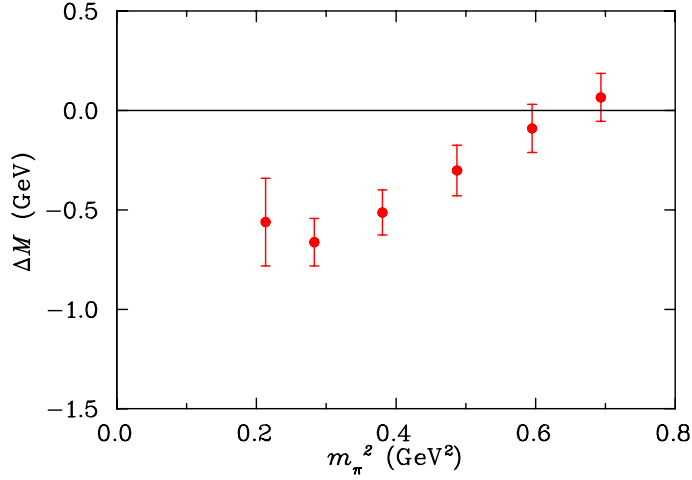} 
\caption{\label{fig:I0_12.pos.s}
	Mass splitting between the $I(J^P)=0(\frac{3}{2}^+)$ state
	extracted with the $NK^{*}$ pentaquark interpolator and the 
	mass of the $P$-wave $N+K$ energy. 
	}
\end{figure}

These $\chi^{2}_{\rm dof}$ analyses have been repeated for all six
quark masses considered with similar results.
Figure~\ref{fig:I0_12.pos.s} shows the mass splitting between the isoscalar
spin-$\frac{3}{2}$ state extracted with the $NK^{*}$ interpolator
and the $P$-wave $N+K$ energy, where we fit at time slices $19-24$ as
concluded above.  Here we are able to show the mass
splitting at all of our six quark masses.  The reason that we are
able to recover a mass splitting for an additional two lighter quark
masses is because, as discussed in the introductory section,
correlated errors in the correlation functions are suppressed in
constructing the effective mass splitting.  The state extracted with
the pentaquark operator indicates the possibility of
binding for the four smallest quark masses shown.

\begin{figure}[tp]
\includegraphics[height=9.0cm,angle=90]{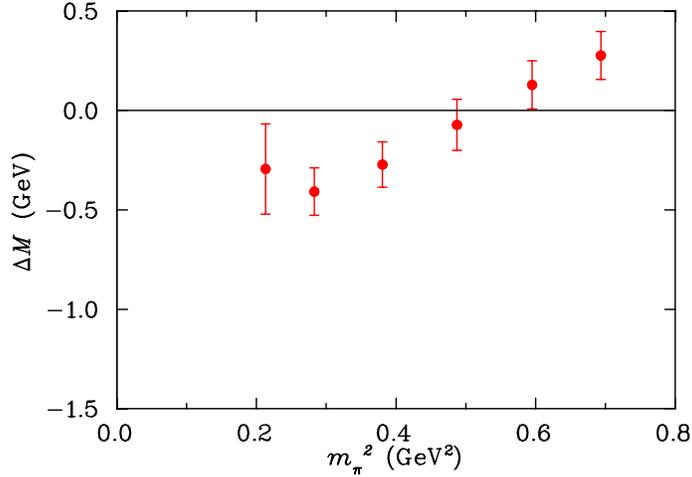} 
\caption{\label{fig:I0_12.pos.s.nk}
	Mass splitting between $I(J^P)=0(\frac{3}{2}^+)$ state
	extracted with the $NK^{*}$ pentaquark interpolator and the 
	two-particle $S$-wave $N+K$ mass threshold. 
}
\end{figure}

In the infinite-volume limit, the lowest energy two-particle state,
the $P$-wave $N+K$, approaches the energy of the $S$-wave $N+K$ state.
As a genuine single-particle state is expected to have a small volume
dependence on our lattice, we also show in Fig.~\ref{fig:I0_12.pos.s.nk}
the mass splitting with the $N+K$ two-particle threshold.
In this figure, the mass difference is also negative, as is
necessary for the presence of a bound state,
suggesting that the presence of binding may prevail on larger
lattices.   Finally, we emphasise the possibility that the mass
splitting may decrease as the light quark mass regime is approached,
allowing the transition to a resonance at physical quark masses.
High statistics studies at lighter quark masses would obviously be of considerable
interest.

\subsection{Negative parity isovector channel}

\begin{figure}[tp]
\includegraphics[height=9.0cm,angle=90]{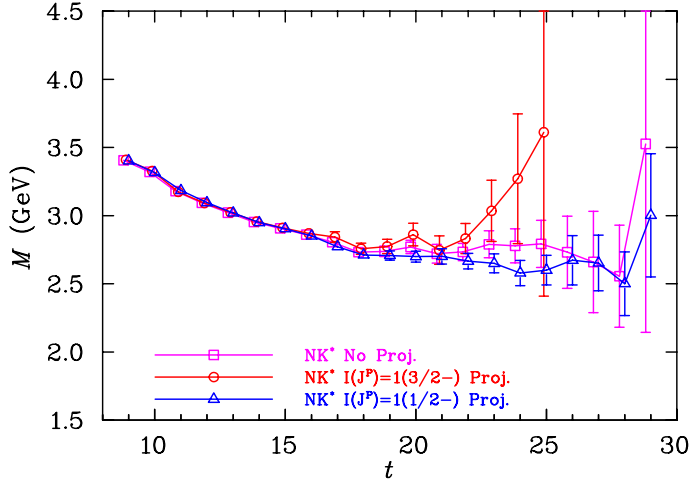} 
\caption{\label{fig:I1.neg_diff}
        As in Fig.~\ref{fig:I0.neg_diff}, but for the isovector
        negative parity channel. }
\end{figure}

\begin{figure}[tp]
\includegraphics[height=9.0cm,angle=90]{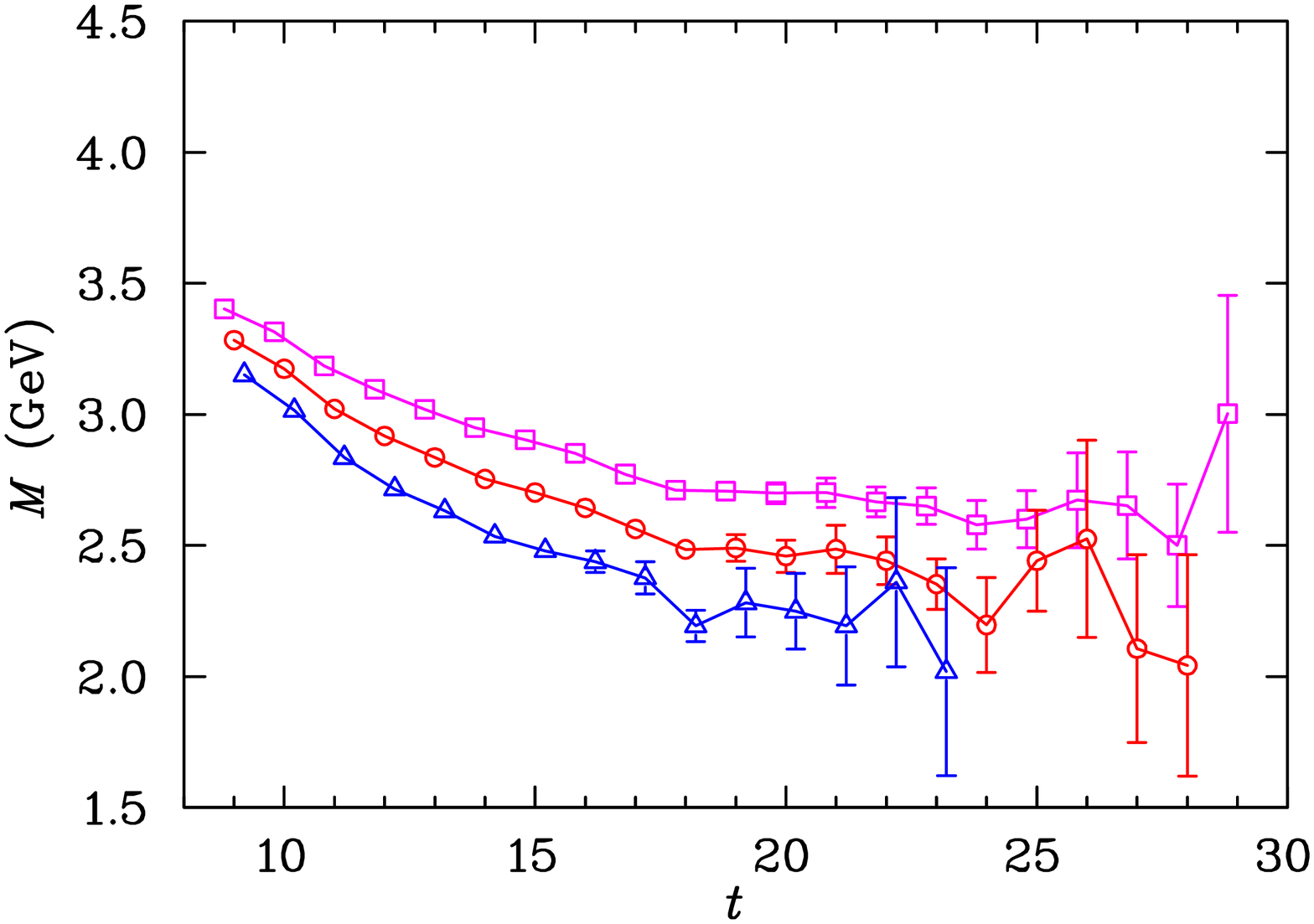} 
\caption{\label{fig:I1.neg_12}
        Effective mass of the $I(J^P)=1(\frac{1}{2}^-)$ pentaquark
	obtained for the $NK^{*}$ interpolator, $\chi_{NK^{*}}$.
	The data correspond to $\mpi\simeq 830{\,{\rm MeV}}$ (squares),
	$700{\,{\rm MeV}}$ (circles), and  $530{\,{\rm MeV}}$ (triangles).}
\end{figure}

\begin{figure}[tp]
\includegraphics[height=9.0cm,angle=90]{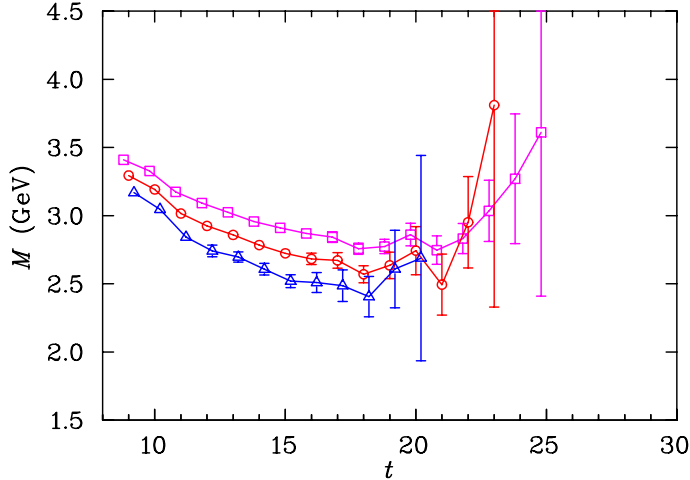} 
\caption{\label{fig:I1.neg_32} 
	As in Fig.~\ref{fig:I1.neg_12}, but for the
	$I(J^P)=1(\frac{3}{2}^-)$ state.}
\end{figure}

For completeness we also include an analysis of the isovector
channel.  First we present the results for the isovector, negative
parity channel.  In Fig.~\ref{fig:I1.neg_diff} we see that the effects
of the spin projection for the largest quark mass are small. This
may be understood by the presence of the $S$-wave $\Delta+K^{*}$ and
$N+K^{*}$ two-particle states, which both the spin-$\frac{1}{2}$ and
spin-$\frac{3}{2}$ projected correlation functions should couple
to. We fit the effective masses extracted from the spin-$\frac{1}{2}$
and spin-$\frac{3}{2}$ projected correlation functions, shown in
Figs.~\ref{fig:I1.neg_12} and \ref{fig:I1.neg_32}, at time slices
$20-30$ and $18-21$, respectively. 

\begin{figure}[tp]
\includegraphics[height=9.0cm,angle=90]{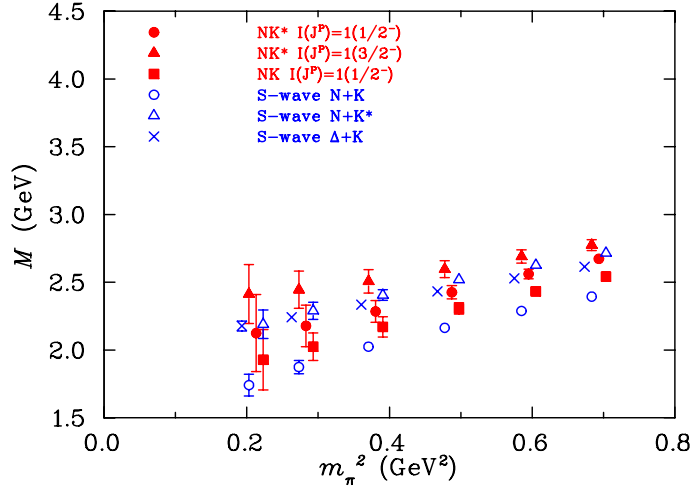} 
\caption{\label{fig:I1.neg}
	Masses of the $I(J^P)=1(\frac{1}{2}^-)$ and
	$1(\frac{3}{2}^-)$ states extracted with the
	$NK^{*}$ interpolating field as a function of
	$m_\pi^2$.  For comparison, we also show the mass of the
	$I(J^P)=1(\frac{1}{2}^-)$ state extracted with the $NK$
	pentaquark interpolator in a correlation-matrix analysis
	\cite{Lasscock:2005tt}, and the masses of the $S$-wave $N+K$,
	$N+K^{*}$ and $\Delta+K$ two-particle states.
	Some of the points have been offset horizontally for clarity.}
\end{figure}

The masses are presented in Fig.~\ref{fig:I1.neg}, along with the
corresponding mass extracted from the $NK$ interpolator and the
relevant two-particle states.  The mass extracted from the
spin-$\frac{3}{2}$ projected correlation function is in good agreement
with the mass of the $N+K^{*}$ two-particle state.  Although 
this correlation function must have a contribution from both the
$\Delta+K$ and $N+K^{*}$ two-particle states, we are probably
accessing an admixture of these states.  A correlation matrix analysis
is required to separate this admixture.  The mass extracted from the
spin-$\frac{1}{2}$ projected correlation function is consistently more
massive in this channel than the mass extracted with the $NK$
interpolator. However, in Ref.~\cite{Lasscock:2005tt} we used a
correlation matrix to extract the mass from the $NK$
interpolator.  This process removes excited state contamination and
renders the ground state mass consistently smaller.  In addition, by
considering the two-point functions calculated with the $NK$ and
$NK^{*}$ interpolators at the quark level, one would naively
expect the $NK^{*}$ interpolator to couple much more
strongly to the $N+K^{*}$ two-particle state than the $NK$
interpolator.  In fact, the weak coupling of the spin-$\frac{3}{2}$
interpolators to the lowest lying spin-$\frac{1}{2}$ states is
reflected in the relatively large statistical uncertainties for the
$NK^{*}$ interpolator results.  

In summary, as in the case of the negative parity isoscalar channel,
there is no indication of the lattice resonance signature.

\subsection{Positive parity isovector channel}

\begin{figure}[tp]
\includegraphics[height=9.0cm,angle=90]{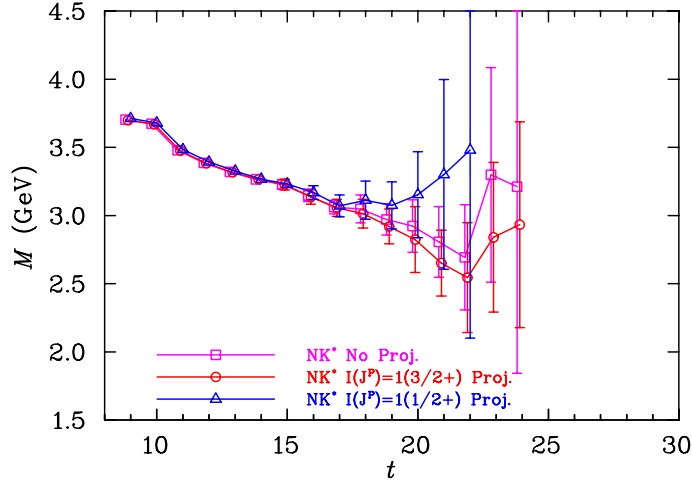} 
\caption{\label{fig:I1.pos_diff}
        As in Fig.~\ref{fig:I0.neg_diff}, but for the isovector
        positive parity channel. }
\end{figure}

\begin{figure}[tp]
\includegraphics[height=9.0cm,angle=90]{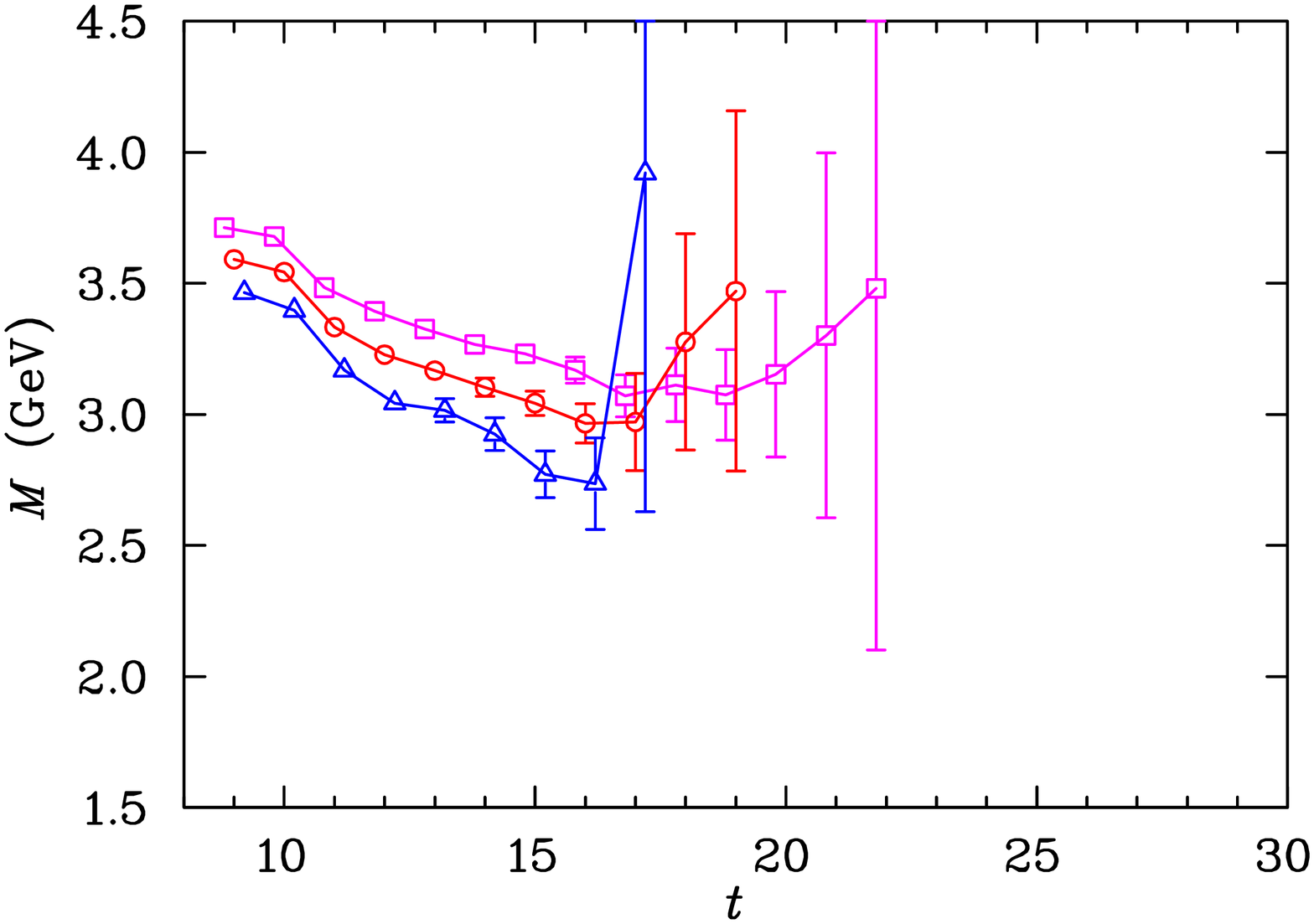} 
\caption{\label{fig:I1.pos_12}
	Effective mass of the $I(J^P)=1(\frac{1}{2}^+)$ pentaquark
	obtained from the $NK^{*}$ interpolator.
	The data correspond to $\mpi\simeq 830{\,{\rm MeV}}$ (squares),
	$700{\,{\rm MeV}}$ (circles), and  $530{\,{\rm MeV}}$ (triangles).}
\end{figure}

\begin{figure}[tp]
\includegraphics[height=9.0cm,angle=90]{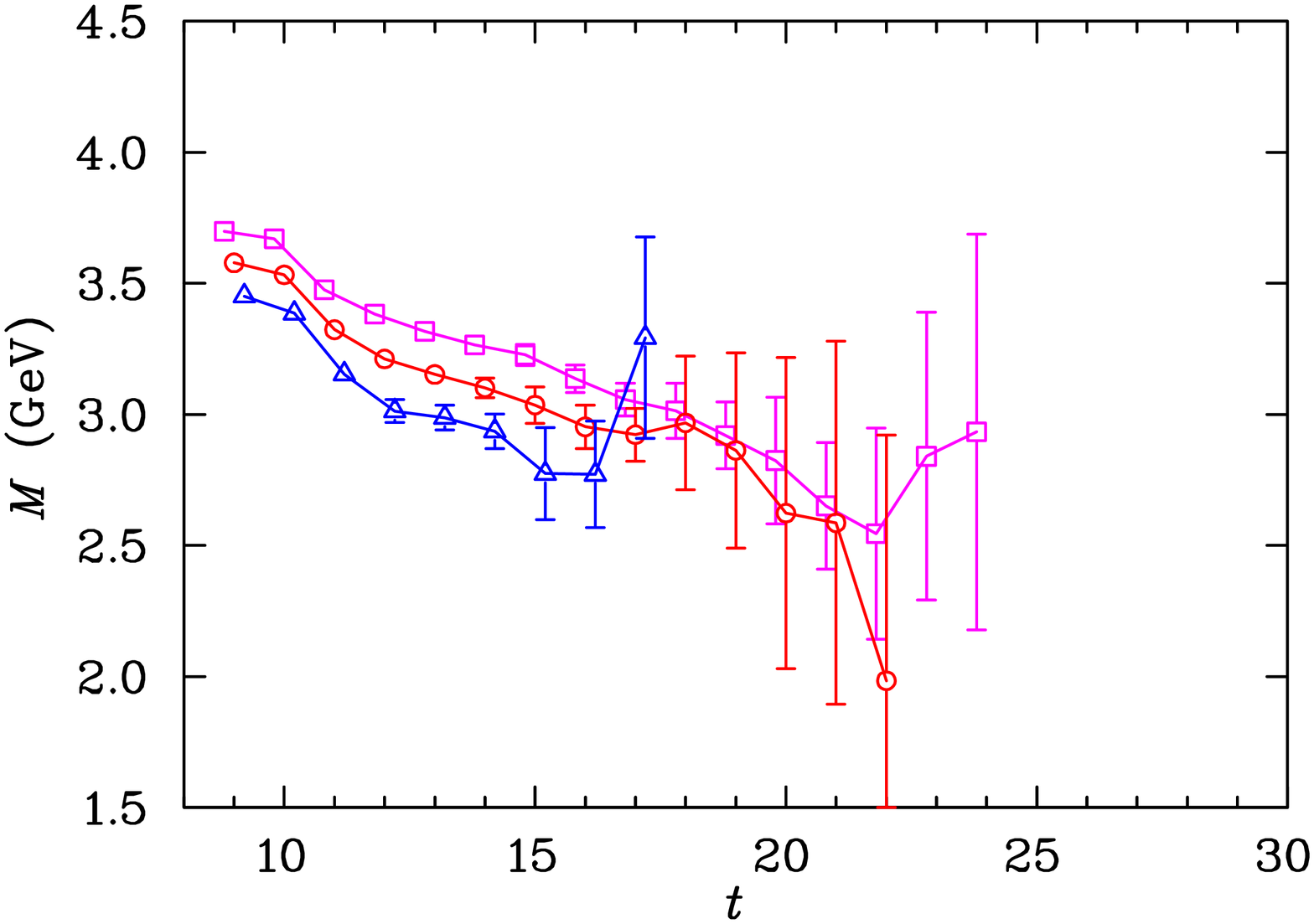} 
\caption{\label{fig:I1.pos_32}
	As in Fig.~\ref{fig:I1.pos_12}, but for the
	$I(J^P)=1(\frac{3}{2}^+)$ state.}
\end{figure}

\begin{figure}[tp]
\includegraphics[height=9.0cm,angle=90]{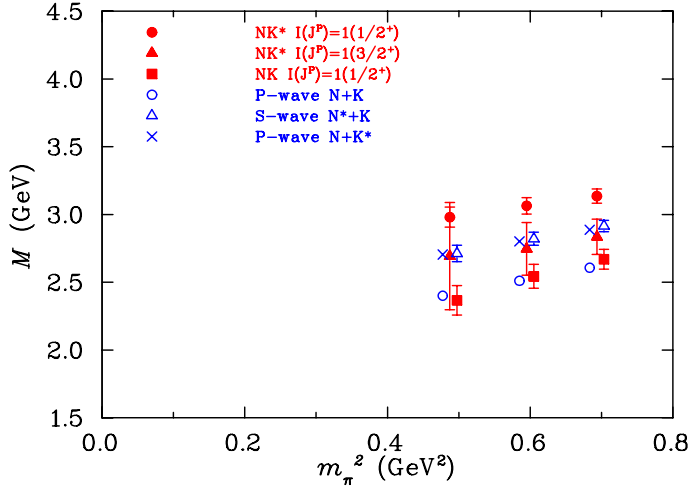} 
\caption{\label{fig:I1.pos}
	Masses of the $I(J^P)=1(\frac{1}{2}^+)$ and
	$1(\frac{3}{2}^+)$ states extracted with the
	$NK^{*}$ interpolating field as a function of
	$m_\pi^2$.  For comparison, we also show the mass of the
	$I(J^P)=1(\frac{1}{2}^+)$ state extracted with the $NK$
	pentaquark interpolator \cite{Lasscock:2005tt}, and the masses
	of the $P$-wave $N+K$, $N+K^{*}$ and the $S$-wave $N^{*}+K$
	two-particle states.  Some of the points have
	been offset horizontally for clarity.}
\end{figure}

Finally we complete our discussion with the positive parity isovector
channel. In Fig.~\ref{fig:I1.pos_diff} we see that the spin projection
suggests the presence of two distinct states but the errors overlap.
The effective masses corresponding to these correlation
functions are presented in Figs.~\ref{fig:I1.pos_12} and
\ref{fig:I1.pos_32}. We fit the effective masses calculated from the
spin-$\frac{1}{2}$ and spin-$\frac{3}{2}$ projected correlation
functions at time slices $19-22$ and $15-20$, respectively.
Due to the poor signal, results for only the three largest quark
masses are shown.  The masses of these states are presented in
Fig.~\ref{fig:I1.pos} along with the mass extracted with the $NK$
interpolator and energies of the relevant two-particle states.  Neither
state extracted with the $NK^{*}$ interpolator
 lies below the lowest energy scattering states, which is necessary
for binding.
Therefore the analogous lattice resonance signature is also absent in this channel.

\section{ Conclusions }

We have completed a comprehensive analysis of the isospin and parity
states of the spin-$\frac{3}{2}$ pentaquark.  Following our previous
work \cite{Lasscock:2005tt}, we search for the standard signature of a
resonance in lattice QCD, where the presence of attraction
renders the resonance mass lower than
the sum of the free decay channel masses
at quark masses near the physical regime.
This standard lattice
resonance signature has been observed for every conventional baryon
resonance ever calculated on the lattice
\cite{Leinweber:2004it,Melnitchouk:2002eg,Zanotti:2003fx,Sasaki:2001nf,Gockeler:2001db}.

In the case of a pentaquark resonance, the relative mass splitting is expected
to vanish in the heavy quark limit.  Therefore the analogous lattice resonance
signature for a pentaquark state will exhibit a negative mass
splitting at intermediate quark masses, with a general trend towards
zero as the heavy quark limit is approached.

In our examination of spin-$\frac{3}{2}$ pentaquark states we have
discovered evidence of the standard lattice resonance signature, in
the spin-$\frac{3}{2}$ positive-parity isoscalar channel.  At
intermediate quark masses, the presence of attraction between the
constituents of the pentaquark baryon is sufficient to render the mass
of the pentaquark state lower than the $N+K$ two-particle threshold.
The mass splitting approaches zero as the light quark masses become
very large, in accord with expectations.  Moreover,
Fig.~\ref{fig:I0_12.pos.s.nk} suggests that the resonance signature
might prevail on larger lattices, and provides a hint that the mass
splitting will decrease as the light quark mass regime is approached,
allowing the transition to a resonance at physical quark masses.

Future work must explore the volume dependence of the binding observed
in the spin-$\frac{3}{2}$ positive-parity isoscalar channel.
Otherwise, one cannot completely rule out the possibility that the
observed binding is a pure finite-volume effect, reflecting a
non-trivial scattering phase shift \cite{Luscher:1986pf} in the
$I(J^P)=0({\frac{3}{2}}^+)$ $N\, K$ scattering channel at momentum $p
= 2 \pi/L$.  If this is the case, the mass splitting will go to zero
as the volume of the lattice is increased.  On the other hand, a
genuine single-particle state is expected to have a small volume
dependence such that the negative mass splitting is preserved in the
infinite volume limit.  As Fig.~\ref{fig:I0_12.pos.s.nk} illustrates,
the observed mass splitting with the $N+K$ two-particle threshold is
sufficient to maintain the lattice resonance signature in the event
that the volume dependence of the spin-$\frac{3}{2}$ positive-parity
isoscalar state is indeed small.

The $I(J^{P})=0(\frac{3}{2}^{+})$ channel is therefore an interesting
pentaquark resonance candidate for further study.  Chiral fermions
such as the overlap fermion action \cite{Narayanan:1993ss} allowing
access to the lightest quark masses \cite{Kamleh:2001ff,Kamleh:2004aw}
should be brought to bear on this particular channel.  High statistics
studies will be vital in rendering a conclusive result.
Alternative resonance signatures such as the volume dependence of the
residue \cite{Mathur:2004jr,Alexandrou:2005gc} or invariance under
hybrid boundary conditions \cite{Ishii:2004qe,Ishii:2004ib} should
also be brought to bear on this most promising channel.
Ultimately, it will be important to investigate the nature of this
state in full QCD, where the dynamical generation of quark-antiquark
pairs is accounted for in the construction of the gauge field
ensemble.

\acknowledgments

We have benefited from helpful discussions with J.~Dudek and K.~Maltman.
We thank the Australian Partnership for Advanced Computing (APAC) and
the Australian National Computing Facility for Lattice Gauge Theory
managed by the South Australian Partnership for Advanced Computing
(SAPAC) for generous grants of supercomputer time which have enabled
this project.  This work was supported by the Australian Research
Council, and the U.S. Department of Energy contract
\mbox{DE-AC05-84ER40150}, under which the Southeastern Universities
Research Association (SURA) operates the Thomas Jefferson National
Accelerator Facility (Jefferson Lab).

%\bibliographystyle{apsrev} 
%\bibliography{bib}

\end{document}